\documentclass[twocolumn,showpacs,showkeys,preprintnumbers,amsmath,amssymb]{revtex4}
\usepackage{graphicx}% Include figure files
\usepackage{dcolumn}% Align table columns on decimal point
\usepackage{bm}% bold math

\begin{document}

%\preprint{APS/123-QED}

\title{Four variants of theory of the second order phase transitions}
% Force line breaks with \\

\author{Leonid S.Metlov}
\email{lsmet@fti.dn.ua}
\affiliation{Donetsk Institute of Physics and Engineering, Ukrainian
Academy of Sciences,
\\83114, R.Luxemburg str. 72, Donetsk, Ukraine}

\thanks{I thank kind my consultant Alexander Elvinovich Filippov for helphool with preparation of paper}

\date{\today}% It is always \today, today,

\begin{abstract}
Because of one-valued connection between the configurational entropy and the order parameter it is possible 
to present the theory of the second order phase transitions in terms of the configurational entropy. 
It is offered a variant of theory, in which the Nernst theorem is obeyed. Within the framework of heterogeneous 
model the phenomena of growth of level of fluctuations and their correlations are analyzed at transition of 
critical point as competitions of kinetic and relaxation processes in the conditions of proximity of two critical points.
\end{abstract}


%Use showkeys class option if keyword
\maketitle

\section{Introduction}
The Landau theory of second order phase transitions (PT-2) was offered in the middles of previous century 
\cite{l37a,l37b}, but interest to it does not weaken to the present, for example, in phase fields theories 
\cite{akv00,esrc02,lpl02,rbkd04,gptwd05,akegny06,rjm09,s06,cn12}. 
By more late researches in the theory of PT-2 very important one-valued connection was set up between 
the order parameter ($OP$) and the configurational entropy \cite{pp79}. Such connection allows choosing as an independent 
variable one of them. PT-2 was based traditionally on the use of $OP$, but a variant will be first considered here, 
in basis of which as an independent variable configurational entropy is fixed.

\section{Connection between the free and internal energy}

Let's set the free energy functional $F\{\varphi(x)\}$ for the non-equilibrium state of the system with given 
$OP$ $\varphi(x)$ in a form typical for PT-2 \cite{pp79}
  \begin{equation}\label{b1}
F\{\varphi(x)\}=F_{0}+\frac{1}{2}\int[c(\triangledown\varphi)^{2}+a\varphi^{2}+\frac{b}{2}\varphi^{4}-2\varphi h]dV.
  \end{equation}

Free energy is here presented, actually, as expansion of a functional $F\{\varphi(x)\}$ in a series over the small
$\varphi(x)$ and its spatial derivatives. The first term in square brackets is energy of the heterogeneous distribution, 
$h(x)$ is the external field. The type of dependence on the spatial derivate $\varphi(x)$ is dictated by considering 
of a homogeneous and isotropic system. In obedience to ideology of PT-2 the free energy depends on a temperature, however 
its dependence on a temperature is concentrated only in a coefficient $a$, which besides changes a sign in a critical 
point. It is considered that other coefficients do not depend on a temperature at all.

For a homogeneous case
  \begin{equation}\label{b2}
f\{\varphi\}=f_{0}+a\varphi^{2}+\frac{b}{2}\varphi^{4}-2\varphi h.
  \end{equation}

Here and below the extensive thermodynamic variables are designated by large characters $F$, $S$ et cetera, 
and their densities are designated by small characters  $f$, $s$ et cetera. 
  
A derivative of the free energy on the temperature differs from entropy by a sign only. 
As the coefficient $a$ depends on temperature only, than differentiating (\ref{b2}), we get \cite{pp79}  
  \begin{equation}\label{b3}
s=-\frac{\alpha}{2T_{c}}\varphi^{2}<0,
  \end{equation}
where $s$ is entropy density, $T_{c}$ is a critical temperature, the constant $\alpha$ does not depend on a temperature, 
and is determined by a relation
  \begin{equation}\label{b4}
a=\alpha\frac{T-T_{c}}{T_{c}},
  \end{equation}
where $T$ is the absolute temperature (thermostate).

Actually, $s$ is not total entropy, but its configurational part only, as pure thermal effects in PT-2 are not 
explicitly considered. In addition, the relation (\ref{b3}) is not general, but a model. However within the framework 
of this model a one-valued connection between $OP$ and the configurational entropy is established. It means that it is 
possible to choose one of these variables as an independent thermodynamic variable and to outline a theory of PT-2, 
for example, not in terms of $OP$, but in terms of configurational entropy. Besides, it prompts the idea for application 
of similar model relations in more wide area (not only for PT-2), including modeling of severe plastic deformation processes.

We mark that a similar situation arises up in the theory of vacancies. There Boltzmann offers a formula, uniquely 
relating the configurational entropy and the vacancy concentration that also allows choosing one of them as an independent 
variable \cite{m11}. It is possible to conclude from it, that all three variables, configurational entropy, defect concentration 
and order parameter, in a different form characterize the same structural reality of solid, its defectiveness. In the case 
of phase transitions the defectiveness obviously can be related to the spontaneous origin of embryos of a new phase.

Within the framework of this model relation it is possible to pass in accordance with expression
  \begin{equation}\label{b5}
F_{c}=U-TS_{c}
  \end{equation}
from the configurational free energy to the internal energy and vice versa. Here, to underline configurational 
nature of free energy and entropy, they are supplied a lower index $c$.

Differentiating (\ref{b5}) we get useful relations
  \begin{equation}\label{b6}
S_{c}=-\frac{\partial F_{c}}{\partial T}=-\frac{\partial U}{\partial T}+S_{c}+T\frac{\partial S_{c}}{\partial T},
  \end{equation}
from which follows
  \begin{equation}\label{b7}
\frac{\partial U}{\partial T}=T\frac{\partial S_{c}}{\partial T}\equiv0
  \end{equation}
by virtue of that the derivative of configurational entropy on temperature is the second derivative of the free 
energy on temperature, but the last depends linearly on the temperature. It is follows from there that both the 
configurational entropy and the internal energy do not explicitly depend on temperature.

Using Eqs. (\ref{b2}) and (\ref{b5}) we find an explicit expression for the internal energy (for a homogeneous case 
and without the account of the external field)
  \begin{equation}\label{b8}
u\{\varphi\}=f_{0}-\frac{\alpha}{2}\varphi^{2}+\frac{b}{4}\varphi^{4},
  \end{equation}
where already all coefficients do not depend on a temperature.

It is possible to consider by virtue of generality of result that namely expression (\ref{b8}) is base model 
relation of theory of PT-2, and temperature dependence of the coefficient $a$ (\ref{b4}) and the free energy 
(\ref{b1}), (\ref{b2}) is simple consequence of this fact.

Indeed, let us consider that simple base relation for internal energy (\ref{b8}), in which all of coefficients 
do not depend on a temperature, is initially given. In this case  $\partial U/\partial T\equiv0$, and according 
to (\ref{b7}) $\partial S_{c} /\partial T\equiv0$ too. Then configurational entropy can be also presented as a 
series expansion on $OP$, limited here quadratic approaching only
  \begin{equation}\label{b9}
s_{c}=c\varphi^{2}+...,
  \end{equation}

Substituting it in Eq. (\ref{b5}) and taking into account Eq. (\ref{b8}) we get
  \begin{equation}\label{b10}
f\{\varphi\}=f_{0}-\alpha\varphi^{2}-cT\varphi^{2}+\frac{b}{2}\varphi^{4}.
  \end{equation}

We define coefficient $c$ from a condition that in the critical point $T=T_{c}$ a total coefficient at $\varphi^{2}$ 
must be a zero
  \begin{equation}\label{b11}
c=-\frac{\alpha}{2T_{c}}.
  \end{equation}

Collecting all formulas, we get (\ref{b2}).

\begin{figure}
\includegraphics [width=3.3 in]{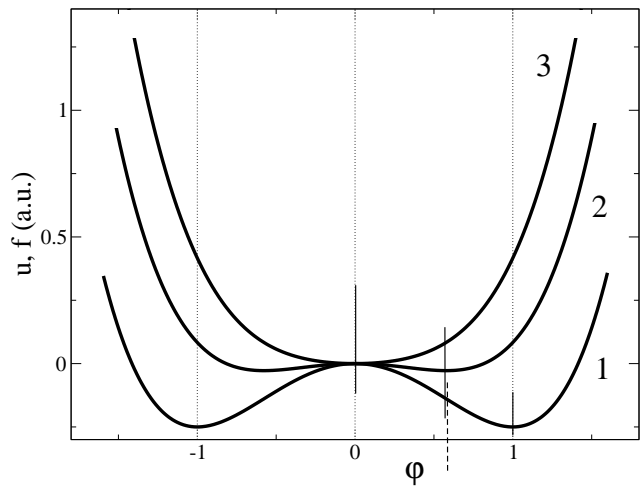}% Here is how to import EPS art
\caption{\label{f1} Internal (1) and free (1-3) energies: 1 -- at $T=0 K$; 2 -- at $T=200 K$; 
3 -- at $T=400 K$. It is here accepted $\alpha=1$, $b=1$, $T_{c}=300 K$.}
\end{figure}

We mark that in area of small $\varphi$ in (\ref{b8}) the first term prevails, and the internal energy has a maximum. 
Therefore for the nonequilibrium states with a zero value $OP$ ($T>T_{c}$) the internal energy has a maximum too 
(curve 1, fig. \ref{f1}), 
while the free energy has, as it must be, a minimum (curve 3). At large $OP$, the second term 
prevails, and the internal energy has already minimums in non-zero extreme points, determined from a conditions
  \begin{equation}\label{b12}
\mu\equiv\frac{\partial u}{\partial \varphi}=\varphi(-\alpha+b\varphi^{2})=0,
  \end{equation}
and equal
  \begin{equation}\label{b13}
\varphi_{\mu}=\pm(\frac{\alpha}{b})^{1/2}.
  \end{equation}

Here $\mu$ is the \textquoteleft\textquoteleft chemical potential\textquoteright\textquoteright as surplus energy, being on unit of $OP$.
  
In area of zero $OP$ the internal energy is convex; it is concave in area of non-zero values of $OP$ (\ref{b13}). 
Inflection points, dividing these areas, are deduced from a condition  
  \begin{equation}\label{b14}
\frac{\partial^{2}u}{\partial \varphi^{2}}=-\alpha+3b\varphi^{2},
  \end{equation}  
 that gives a value 
  \begin{equation}\label{b15}
\varphi_{\mu}=\pm(\frac{\alpha}{3b})^{1/2}
  \end{equation}
(dotted vertical line in fig.1).

Now we compare positions of extremums of the internal and free energy. For the last they are determined from a condition
  \begin{equation}\label{b16}
\frac{\partial f}{\partial \varphi}=\varphi(a+b\varphi^{2})=0,
  \end{equation}
and non-zero roots are equal
  \begin{equation}\label{b17}
\varphi_{F}=\pm(-\frac{a}{b})^{1/2}=\varphi_{\mu}(\frac{T_{c}-T}{T_{c}})^{1/2}
  \end{equation}

From where we notice that at a zero temperature non-zero roots of the free and internal energy coincide 
(their graphs coincide fully, see curve 1 in fig. \ref{f1})
  \begin{equation}\label{b18}
\varphi_{\mu}=max(\varphi_{F})=\pm(\frac{\alpha}{b})^{1/2}\neq 0.
  \end{equation}

With growth of temperature non-zero roots of the free energy diminish (curve 2) and in a critical point 
goes to zero (curve 3). Here they meet with identically zero roots. Higher than critical point expression 
(\ref{b17}) is lost meaning, and there are only roots of equations (\ref{b16}) identically equal to the zero.
  
\begin{figure}
\includegraphics [width=3.3 in]{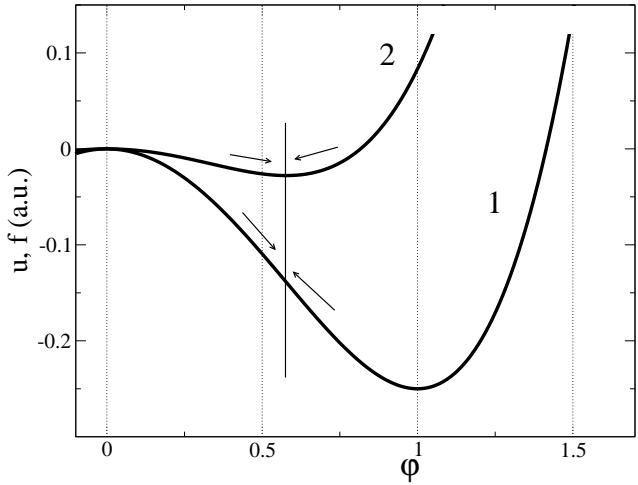}% Here is how to import EPS art
\caption{\label{f2} Internal (1) and free (2) energies at $T=200 K$, Arrows pointer specify direction 
of the system evolution along relief of the free and internal energy.}
\end{figure}

If the system is in the non-equilibrium state, it tends to the equilibrium state in accordance with 
the Landau –- Khalatnikov equation (see arrows along to the curve 2, fig. \ref{f2}).  
  \begin{equation}\label{b19}
\frac{\partial \varphi}{\partial t}=-\gamma_{f}\frac{\partial f}{\partial \varphi},
  \end{equation}  
where $f$ is the free energy density, $\gamma_{f}$ is a kinetic coefficient.
  
\begin{figure}
\includegraphics [width=3.3 in]{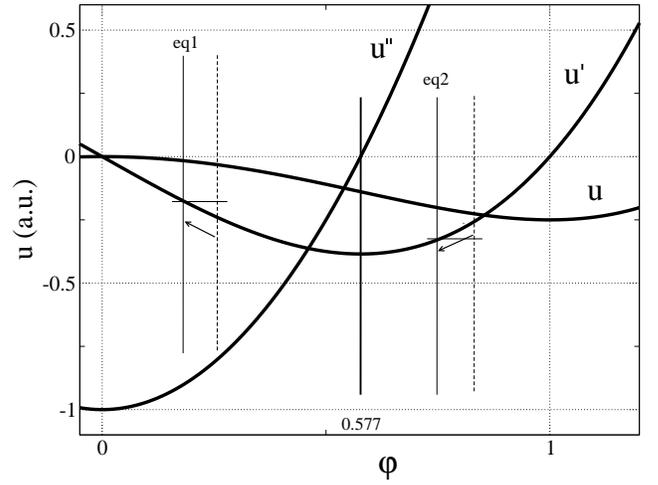}% Here is how to import EPS art
\caption{\label{f3} Internal energy and its derivatives. Vertical line is conducted through minimums 
of free energy ($eq1$, $eq2)$}
\end{figure}

But tending of the system to the equilibrium state it is possible to express and in terms of the internal energy. 
We must take into account thus that in the equilibrium state the tangent to the graph of the free energy has 
a zero inclination by definition, while tangent to the graph of the internal energy has a non-zero inclination. 
Then evolution equation in terms of the internal energy must look like   
  \begin{equation}\label{b20}
\frac{\partial \varphi}{\partial t}=\pm\gamma_{u}(\frac{\partial u}{\partial \varphi}-\mu_{eq}),
  \end{equation}  
where $u$ is the internal energy density, $\gamma_{u}$ is a new kinetic coefficient, $\mu_{eq}$ is 
\textquoteleft\textquoteleft chemical potential\textquoteright\textquoteright in the equilibrium state. 
Sign a \textquoteleft\textquoteleft plus\textquoteright\textquoteright gets out, if in the equilibrium state 
the internal energy is convex, sign \textquoteleft\textquoteleft minus\textquoteright\textquoteright – if concave 
(fig. \ref{f3}). 
Tending of the system to the equilibrium state in this case is shown by arrows along relief of the internal energy 
in fig. \ref{f2}, \ref{f3}. Both in terms of the free energy and in terms of the internal energy, the system tends 
to the same steady-state (fig. \ref{f2}). It follows from that in a steady-state a condition of equality to zero of 
right parts of evolution Eqs. (\ref{b19}) and (\ref{b20}) is satisfied at the same agreed solutions (\ref{b13}) and 
(\ref{b17}).  

Eq. (\ref{b20}) can be directly deduced from Eq. (\ref{b19}). For this purpose it is enough to substitute (\ref{b5}) 
in (\ref{b19}) with taking in account (\ref{b3})  
  \begin{equation}\label{b21}
\frac{\partial \varphi}{\partial t}=-\gamma_{f}(\frac{\partial u}{\partial \varphi}-T\frac{\partial s}{\partial \varphi})
=-\gamma_{f}(\frac{\partial u}{\partial \varphi}-\alpha\frac{T}{T_{c}}\varphi),
  \end{equation}  
  
We consider that deviation from the equilibrium state is small, and value of the second term in (\ref{b21}) 
is little differ from equilibrium. Then, taking the equilibrium value $OP$ from (\ref{b17}), we get 
  \begin{equation}\label{b22}
\frac{\partial \varphi}{\partial t}=-\gamma_{f}(\frac{\partial u}{\partial \varphi}-\mu_{eq}),
  \end{equation}  
where
  \begin{equation}\label{b23}
\mu_{eq}=-\frac{\alpha^{1/2}}{b^{3/2}}\frac{T}{T_{c}}\sqrt{\frac{T_{c}-T}{T_{c}}}.
  \end{equation}  
  
\begin{figure}
\includegraphics [width=3.3 in]{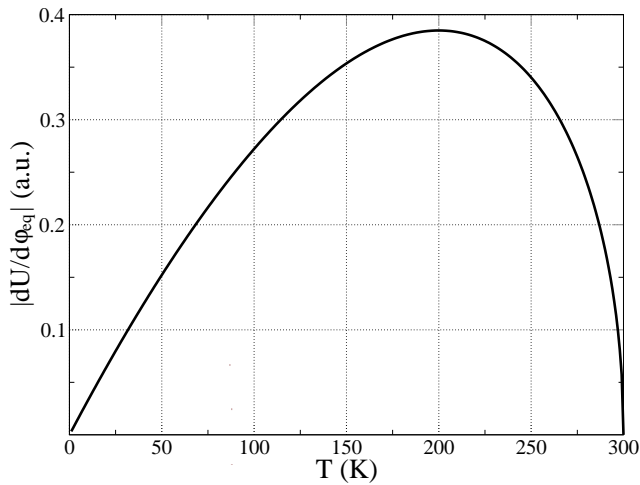}% Here is how to import EPS art
\caption{\label{f4} Dependence of the ``camical potential'' on the temperature}
\end{figure}

Expression for $\mu_{eq}$ it is possible to get also, substituting (\ref{b17}) in (\ref{b12}). 
The curve of its dependence on a temperature is resulted in fig. \ref{f4}, 
from which is obvious, that $\mu_{eq}$ always less zero, that however is obvious and from the curve $U'$ 
in fig. \ref{f3}. In addition, from fig. \ref{f4} it is obvious that most sharply $\mu_{eq}$ changes at 
approaching to the critical point.  
  
Eq. (\ref{b22}) coincides with Eq. (\ref{b20}), if to consider that kinetic coefficients are connected with a relation 
  \begin{equation}\label{b24}
\gamma_{u}=\gamma_{f}sign(u'').
  \end{equation}  
  
The validity of choice of signs in Eq. (\ref{b20}) it is possible to check with help of fig. \ref{f3}, 
where the curves of the internal energy and its first two derivatives on $OP$ are resulted only. 
For illustration the equilibrium states are chosen on the left and on the right of inflection point in area 
of convex and concavity of the internal energy. 

If the system is in the non-equilibrium state $\varphi>\varphi_{eq1}$ (dotted line on the right of $eq1$
in fig. \ref{f3}), then
  \begin{equation}\label{b25}
\frac{\partial u}{\partial \varphi}<(\frac{\partial u}{\partial \varphi})_{eq1},
  \end{equation}
and returning force is directed toward diminishing of $OP$, and the sign in Eq. (\ref{b20}) must be positive. 
In area of concavity at condition $\varphi>\varphi_{eq2}$ inequality is just opposite 
  \begin{equation}\label{b26}
\frac{\partial u}{\partial \varphi}>(\frac{\partial u}{\partial \varphi})_{eq2},
  \end{equation}
and sign in Eq. (\ref{b20}) must be chosen negative.

Thus, following connection between the configurational entropy and $OP$ (\ref{b3}), it was succeeded to set 
connection between the free and internal energy, and to find the alternative form of evolution equations in 
terms of the internal energy (\ref{b20}). In an order to use evolution equation (\ref{b20}) it is needed to 
know an equilibrium value of \textquoteleft\textquoteleft chemical potential\textquoteright\textquoteright. 
For its finding again all the same it is necessary to use 
a minimum of the free energy (\ref{b16}), in which it is possible to calculate all of descriptions of the 
internal energy and its derivatives.

\section{Fluctuation formulation of problem in terms of internal energy}

For a heterogeneous problem a functional of the internal energy by analogy with (\ref{b1}) and with taking 
in account (\ref{b8}) looks like
  \begin{equation}\label{b27}
U\{\varphi(x)\}=F_{0}+\frac{1}{2}\int[c(\triangledown\varphi)^{2}-\alpha\varphi^{2}+\frac{b}{2}\varphi^{4}-2\varphi h]dV.
  \end{equation}

Evolution equation of type (\ref{b20}) is in this case
  \begin{equation}\label{b28}
\frac{\partial \varphi}{\partial t}=\pm\gamma_{u}(\frac{\delta u}{\delta \varphi}-\mu_{eq}),
  \end{equation}
or in an explicit form
  \begin{equation}\label{b29}
\frac{\partial \varphi}{\partial t}=\pm\gamma_{u}(-c\bigtriangleup\varphi-\alpha\varphi+b\varphi^{3}-h-\mu_{eq}),
  \end{equation}

Equation contains algebraic part (sources and sinks) and differential one. If to ignore algebraic part 
at a negative coefficient $c<0$ the equation is diffusive type. In such form it can describe the processes 
of spreading (diffusion) of $OP$, resulting in its more homogeneous distribution and, consequently, it is 
favorable for «resorption» of possible fluctuations. Vice versa, at a positive sign $c>0$ this equation can 
describe the processes of strengthening of fluctuations or avalanche-type transition to the new phase.

«Chemical potential» of $OP$ unlike (\ref{b12}) is now determined through a functional derivative
  \begin{equation}\label{b30}
\mu=\frac{\delta u}{\delta \varphi}=-c\bigtriangleup\varphi-\alpha\varphi+b\varphi^{3}-h,
  \end{equation}
and it depends on gradient part. At the same time, its equilibrium value $\mu_{eq}$ must not depend on gradient part, 
because the equilibrium state is supposed the homogeneous distribution by definition. Therefore an equilibrium value 
is determined on those formulas (\ref{b23}) as for a homogeneous problem.

The evolution equation (\ref{b29}), following from functional of the internal energy (\ref{b27}), as well as 
evolution equations, followings from functional of free energy (\ref{b1}) (see \cite{pp79}) can describe relaxation 
(suppression) of the heterogeneous field of fluctuations. But they do not contain an active constituent, 
describing the generation of thermal fluctuations (nois). For modeling of this we add an accidental source 
of $OP$ to right part (\ref{b29})
  \begin{equation}\label{b31}
\frac{\partial \varphi}{\partial t}=\pm\gamma_{u}(-c\bigtriangleup\varphi-\alpha\varphi+b\varphi^{3}-h-\mu_{eq})
+n(\varphi),
  \end{equation}

Thus, if the system is initially in the equilibrium state and a temperature strongly differs from critical one, 
expression in parentheses in Eq. 31 equals zero for all volume of the system. The origin of thermal fluctuations 
of $OP$ due to the last term transfers locally the system into separate areas in a non-equilibrium state. Now for 
these areas expression in parentheses becomes different from zero, and the reaction of the system is directed on 
suppression of arising up fluctuations. Note that suppression them goes in all of volume due to algebraic part of 
the evolution equation (rapid process), and additionally due to gradient part on the boundaries of areas (slow process).

At the same time, fluctuations are arisen in other places.  The processes of their generation and suppression, 
which will dynamically counterbalance each other, go in parallel. Actually, thermal fluctuations displace slightly 
the true equilibrium state of the system, and it will take nature of stationary-state.

These processes in vicinity of a critical point, when two (zero and non-zero) steady-states are close to each other, 
will go quite othergates. Thermal fluctuation can transfer part of volume of the system from one stable state in another 
stable state. In this case, the volume (rapid) suppression of fluctuation is absent, and there is only slow suppression 
it on the boundaries of area and the boundary of an area will be gradually reduced. As this process is slow by virtue 
of dimension factor, long-living fluctuations are arisen. As the process of generation of fluctuations continues with 
same intensity, and the process of their suppression is strongly slowed, the total number of fluctuations increases, 
what is observed at PT-2.

\begin{figure}
\includegraphics [width=0.79 in]{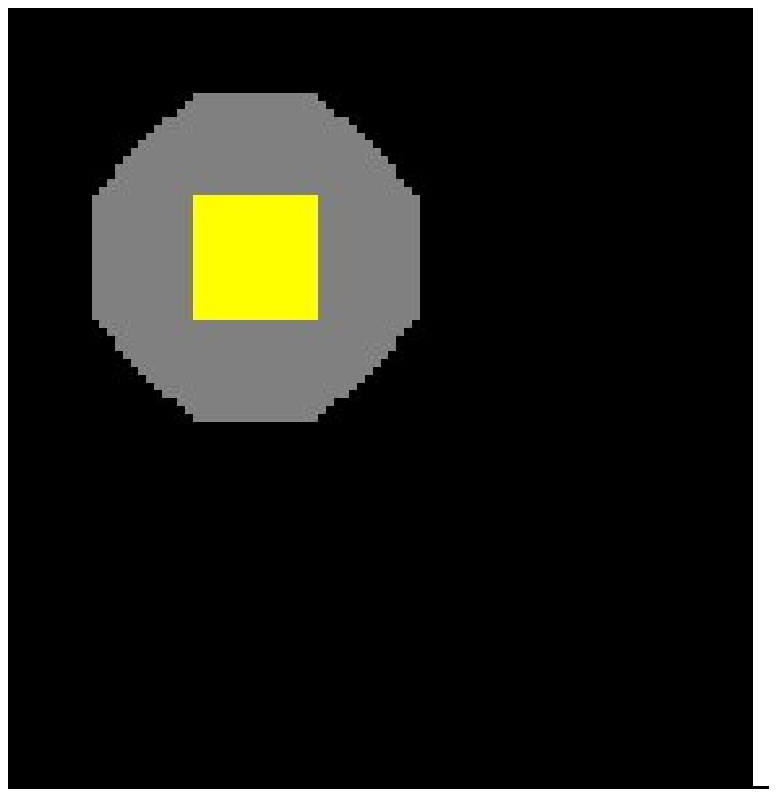}% Here is how to import EPS art
\includegraphics [width=0.79 in]{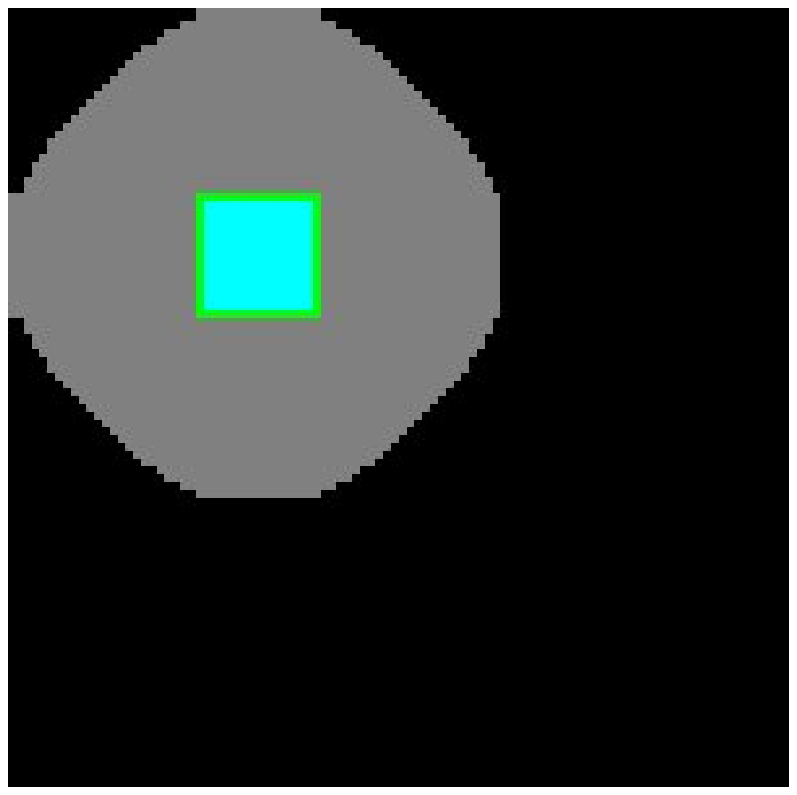}
\includegraphics [width=0.79 in]{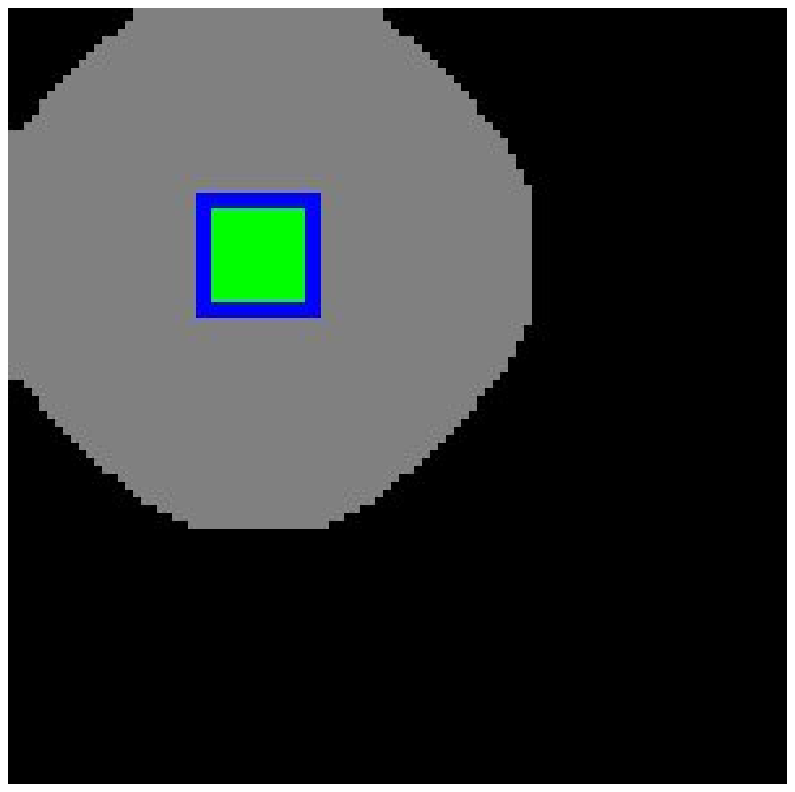}
\includegraphics [width=0.79 in]{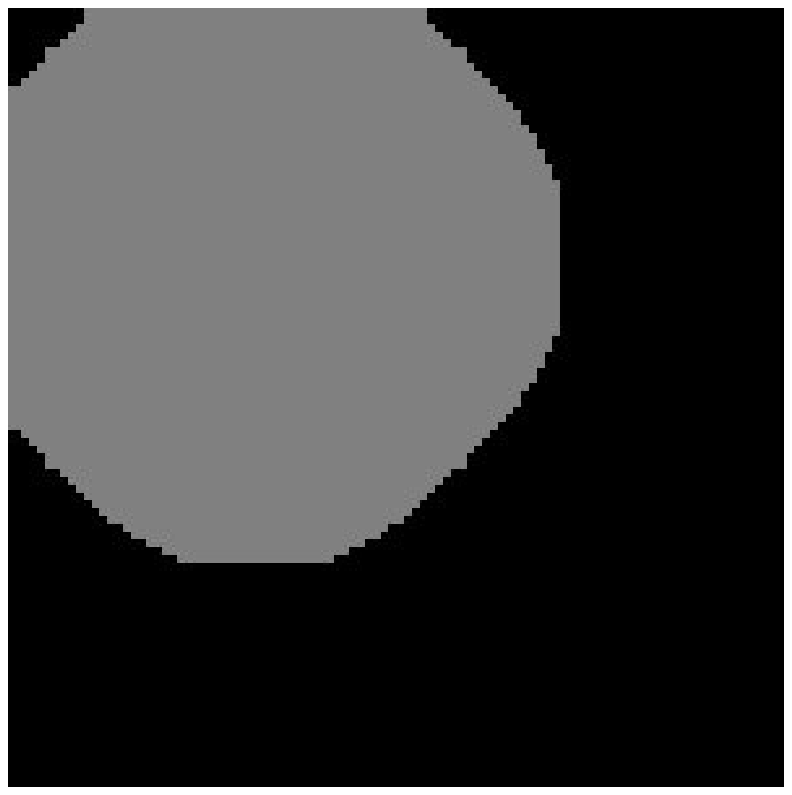}
\caption{\label{f5} Evolution of 2D system with heterogeneity through the equal intervals of time.}
\end{figure}

It is of interest to probe numerically the transition of the system through a critical point on some model example. 
In view of calculable resource limiting of serial computers we consider the $2D$ variant of problem. Parameters 
for calculations the same, as higher (to fig. \ref{f1}), a coefficient at a gradient term is chosen equal 
$c=0.5$. All of area with zero OP consists of $100\times100$ squares of unit sizes. For testing of problem a 
heterogeneity is entered in the left overhead corner of model with co-ordinates $25\div40$ on a horizontal line and 
on a vertical line (fig. \ref{f5}). 
$OP$ in this area gets out equal $0.1$.

The evolution of heterogeneity goes in an expected manner. The area of heterogeneity «diffuses», broadening in size. 
Amplitude diminishes here, both due to diffusion and due to relaxation in a volume. The evolution is slowed in the 
course of time, and the system tends to pass fully to the equilibrium state.

\begin{figure}
\includegraphics [width=0.79 in]{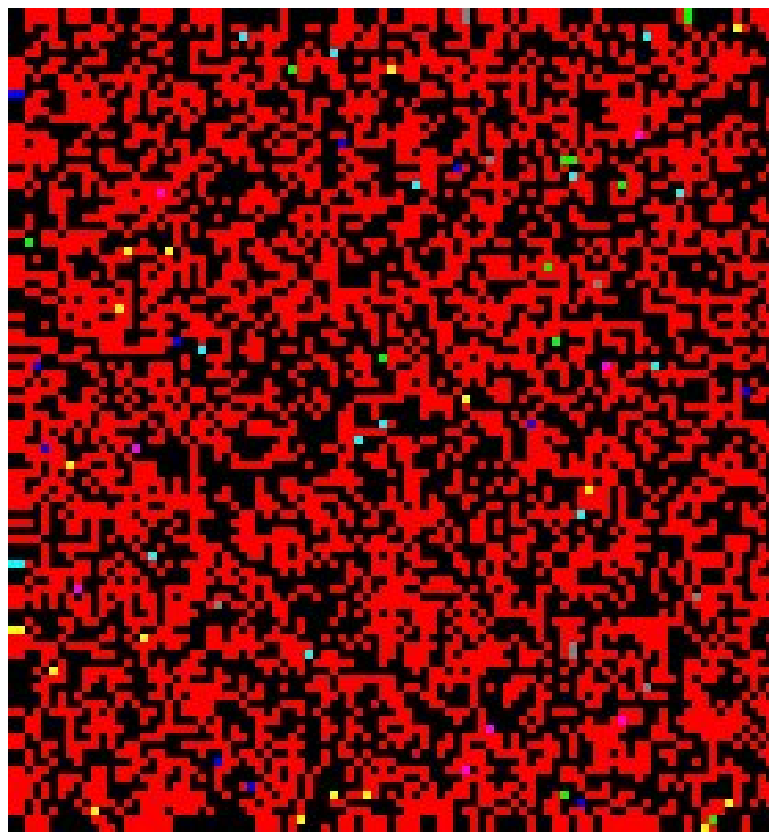}% Here is how to import EPS art
\includegraphics [width=0.79 in]{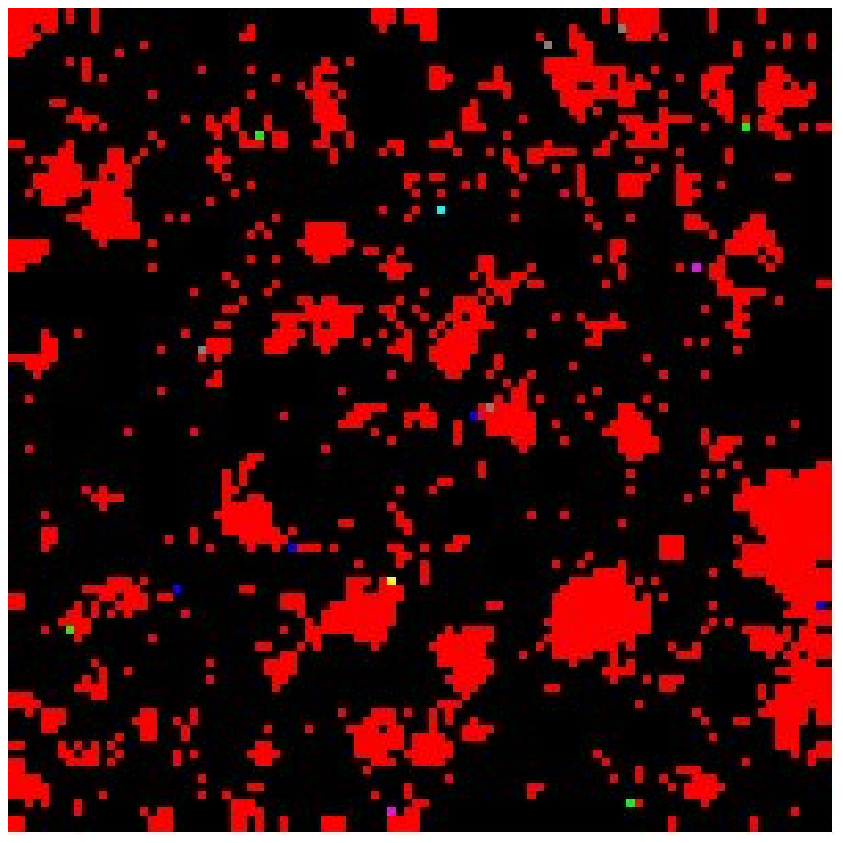}
\includegraphics [width=0.79 in]{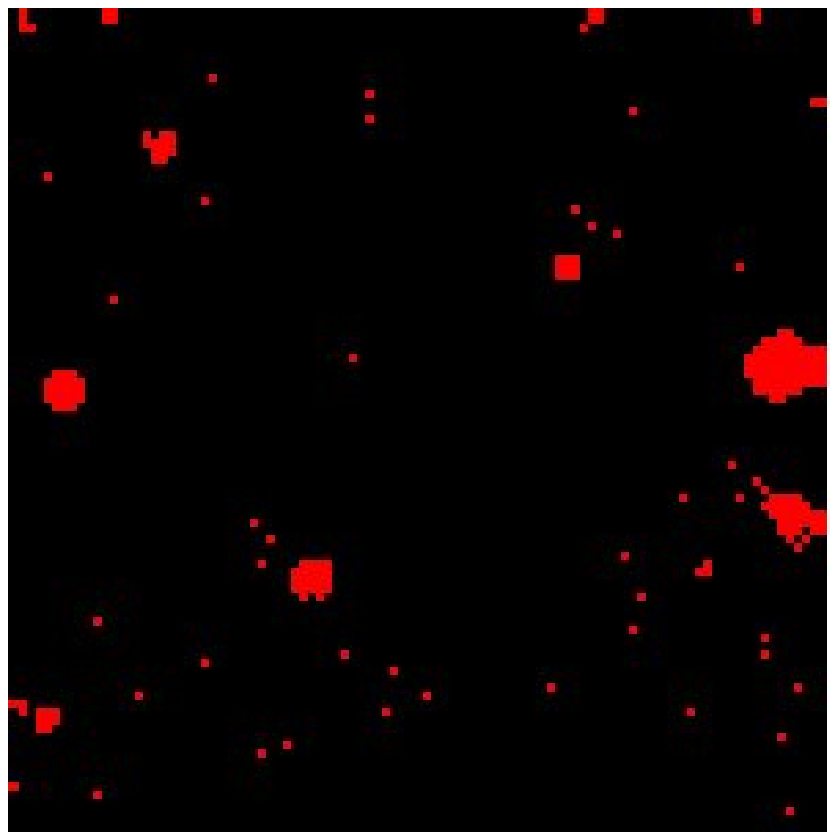}
\includegraphics [width=0.79 in]{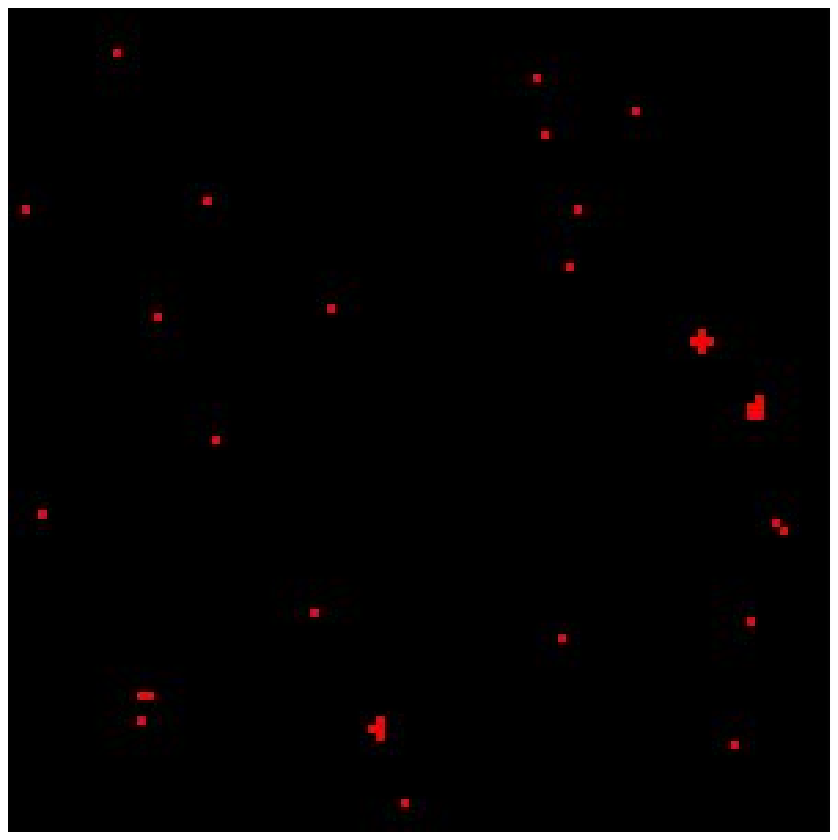}
\caption{\label{f6} Evolution of 2D system with heterogeneity through the equal intervals of time.}
\end{figure}

For modeling of accidental fluctuations the function of sources $n(\varphi)$ in (\ref{b31}) chooses in a form of 
«white noise» with amplitude $0.1$, setting accidentally in every cell. The example of the system evolution is 
resulted in fig. \ref{f6}.

We see that fluctuations in the initial state have a branching fractal-like structure, (the first frame in fig. \ref{f6}). 
On later stages of evolution they are localized in hearth-like areas (the second frame in fig. \ref{f6}), which becomes 
less and less in number, and they diminish in size (the third frame in fig. \ref{f6}). Finally, they disappear 
practically from the visible field (the last frame in fig. \ref{f6}). At the chosen relationship between intensity 
of production of fluctuations and their annihilation, the last prevails. At other parameters other situation is possible.

Unfortunately, a transition through the critical point $T_{c}$ did not give the effect of growing of fluctuations; 
therefore investigation of this case will be executed in the next section, in which the theory of PT-2 is presented 
in terms of the configurational entropy.

\section{Problem of PT-2 in terms of the configurational entropy}

Relation (\ref{b3}) shows a potential possibility for formulation of theory of phase transitions without resorting 
to the concept of order parameter, but being based directly on (configurational) entropy. It is here necessary to 
rewrite the base relation (\ref{b3}) in a form
  \begin{equation}\label{b32}
\varphi=\pm s\sqrt{-\frac{2T_{c}}{\alpha}},
  \end{equation}

Because of that, a sign of expression under a root is «minus», entropy can be either a negative value that corresponds 
partial or complete ordering or identical zero that corresponds the complete disordering. The free energy (\ref{b2}) 
in absence of an external field must be writen down in a form
\begin{eqnarray}\label{b33}
\nonumber
f(s)=f_{0}+2(-(T-T_{c})s+b(\frac{T_{c}}{\alpha})^{2}s^{2})\;at\;T<T_{c}, \\
f(s)=f_{0},\;at\;T>T_{c}. 
\end{eqnarray}

In this formulation, however, the free energy can not be presented by unified expression for all of temperature interval, 
but it is presented as a locally determined function. It is related to that fact that the configurational entropy peaks 
at the temperature of $T=Tc$, and at further growth of temperature it remains at this (zero) value. The equilibrium 
values are determined from a condition
\begin{eqnarray}\label{b34}
\nonumber
\frac{\partial f}{\partial s}=2(-(T-T_{c})+2b(\frac{T_{c}}{\alpha})^{2}s)=0\;at\;T<T_{c}, \\
\frac{\partial f}{\partial s}\equiv0\;at\;T>T_{c}. 
\end{eqnarray}

From where
\begin{eqnarray}\label{b35}
\nonumber
s_{1}=\frac{1}{2b}(T-T_{c})(\frac{\alpha}{T_{c}})^{2}\;at\;T<T_{c}, \\
s_{2}=0\;at\;T>T_{c}. 
\end{eqnarray}

At $T>T_{c}$ the system is in a state of indifferent equilibrium. Any constant $s$ formally satisfies the condition, 
however, from considering of continuity of the free energy and its first derivatives it follows to choose the second 
equilibrium value as zero.

In principle, it is necessary to write separately Landau -- Kalatnikov-like evolution equations for every temperature 
interval. For the interval $T<T_{c}$ the type of equation is quite obvious
  \begin{equation}\label{b36}
\frac{\partial s}{\partial t}=-\gamma_{1}\frac{\partial f}{\partial s}=-2\gamma_{1}(-(T-T_{c})+2b(\frac{T_{c}}{\alpha})^{2}s),
  \end{equation}

For the interval $T>T_{c}$ situation is more difficult. Here the system is in the equilibrium state with a maximal chaos. 
Deviation from this equilibrium state can be only by fluctuation toward diminishing of chaos and appearance of partial 
ordering. This effect can be only small of the second order with respect to entropy and expansion of the free energy is 
begun with the quadratic term on entropy. Therefore we will specify the second equation (\ref{b33})
  \begin{equation}\label{b37}
f(s)=f_{0}+2b(\frac{T_{c}}{\alpha})^{2}s^{2}\;at\;T>T_{c}.
  \end{equation}

A coefficient is chosen from those considering, that relief of the free energy must be continuously changed during 
transition of the critical point. The first Eq. \ref{b33} and Eq. \ref{b37} can be written as unified equation, 
if we use theta-function
  \begin{equation}\label{b38}
f(s)=f_{0}-(T-T_{c})s\varTheta(T_{c}-T)+2b(\frac{T_{c}}{\alpha})^{2}s^{2}+....
  \end{equation}

\begin{figure}
\includegraphics [width=3.3 in]{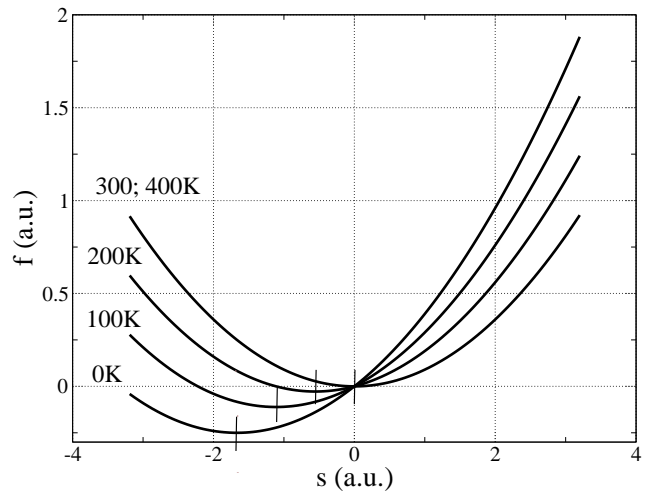}% Here is how to import EPS art
\caption{\label{f7} Dependence of the free energy on the entropy}
\end{figure}

The curve of the free energy is resulted in fig. \ref{f7}. 
From picture evidently that with growth of temperature a minimum of the free energy is uniformly displaced to the right, 
reaching in the critical point $T=T_{c}=300 K$ of a zero value, and relief of the free energy does not change whereupon.

With taking in account Eq. \ref{b37} evolution equation is for the case $T>T_{c}$ looks like
  \begin{equation}\label{b39}
\frac{\partial s}{\partial t}=-\gamma_{1}\frac{\partial f}{\partial s}=-4\gamma_{1}b(\frac{T_{c}}{\alpha})^{2}s,
  \end{equation}
and with use of theta-function both Eq. \ref{b36} and Eq. \ref{b39} can be also written as one equation 
for all temperature interval
  \begin{equation}\label{b40}
\frac{\partial s}{\partial t}=-2\gamma_{1}(-(T-T_{c})\varTheta(T_{c}-T)+2b(\frac{T_{c}}{\alpha})^{2}s),
  \end{equation}

It is of interest to look, to what Landau -- Khalatnikov equation (\ref{b19}) transits at formulation of the problem 
in terms of the configurational entropy
  \begin{equation}\label{b41}
\frac{\partial s}{\partial t}=-\gamma_{f}\frac{\partial f}{\partial s}(\frac{\partial s}{\partial \varphi})^{2},
  \end{equation}
or with taking in account (\ref{b32}) and explicit expression for the free energy (\ref{b33}) and (\ref{b37})
\begin{eqnarray}\label{b42}
\nonumber
\frac{\partial s}{\partial t}=-2\gamma_{f}\frac{\alpha}{T_{c}}s(T-T_{c}-2b(\frac{T_{c}}{\alpha})^{2}s)\;at\;T<T_{c}, \\
\frac{\partial s}{\partial t}=-4\gamma_{f}b(\frac{T_{c}}{\alpha})s^{2}\;at\;T>T_{c}. 
\end{eqnarray}  
  
The first evolution equation in a form (\ref{b42}) absorbs seemingly in itself both evolution equations (\ref{b36}) 
and (\ref{b39}) and thus it can be extended in all temperature interval. Indeed, the multiplier $s$ causes the system 
to tend to the same steady-state as Eq. \ref{b39}, and multiplier in parentheses – to the same steady-state for 
Eq. \ref{b36}. But it is pure outward coincidence, because the multiplier $s$ arose up in this equation as a formal 
transformation of variables with Jacobian $\partial s/\partial \varphi$ in (\ref{b40}), and it does not any relation 
to physics of process. The second Eq. \ref{b41}, which is deprived every sense, testifies about it too, because any 
negative fluctuation of the entropy causes its further decrease to $-\infty$.

Therefore it is most correct to use Eq. \ref{b40} for analysis of evolution of a system in terms of the configurational 
entropy. It does not result by means of limiting transition from the classic Landau -- Khalatnikov equation (\ref{b19}) 
in terms of $OP$. Taking into account that the configurational entropy is more fundamental quantity as compared to $OP$, 
it is necessary to give a preference for it. It is possible to suppose that an attempt to write down the general 
evolution equation (\ref{b19}) with help of $OP$ at once for a temperature higher and below of critical point contains 
a latent defect, which, though does not influence on the asymptotic states of the system, but can distort speed kinetics 
of phase transition.

In accordance with (\ref{b18}) and (\ref{b3}) at the absolute zero of temperature the configurational entropy is 
minimal and negative. It contradicts to the Nernst theorem, in obedience to which entropy at the zero of temperatures 
must be equal to the zero. For the removal of this contradiction it is enough to shift a scale on entropy,
  \begin{equation}\label{b43}
s\prime=s+\frac{\alpha^{2}}{2bT_{c}},
  \end{equation}
that at zero temperature the configurational entropy is wittingly equal zero. Then the free energy is look like
\begin{eqnarray}\label{b44}
\nonumber
f(s)=f_{0}+\frac{\alpha^{2}}{2bT_{c}}(T-T_{c})\varTheta(T_{c}-T)+\frac{\alpha^{2}}{4b}-\\
-(T-T_{c})\varTheta(T_{c}-T)s\prime-T_{c}s\prime+b(\frac{T_{c}}{\alpha})^{2}s\prime^{2}+.... 
\end{eqnarray}

\begin{figure}
\includegraphics [width=3.3 in]{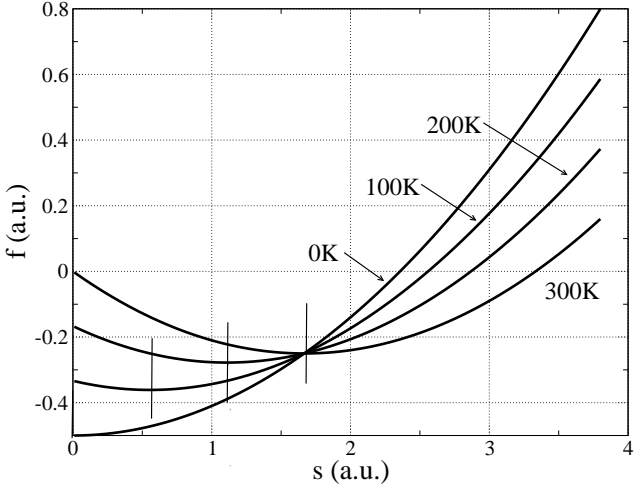}% Here is how to import EPS art
\caption{\label{f8} Dependence of the free energy on the entropy with satisfaction of Nernst theorem}
\end{figure}

At a zero temperature a zero value of the entropy corresponds to a minimum of the free energy, that is, in accordance 
with the Nernst theorem (fig. \ref{f8}). 
With growth of the temperature the entropy grows evenly and takes the maximal value in the critical point 
$T=T_{c}=300 K$. Further with growth of the temperature a relief of the free energy does not change, the system 
reaches a maximal structural disorder, and entropy remains at the attained maximal value. The negative values of 
configurational entropy are throw-away as unphysical. A minimum value of the free energy grows with growth of the 
temperature, as well as in case of fig. \ref{f7}.

Evolution equation (\ref{b40}) for this case is looked like
  \begin{equation}\label{b45}
\frac{\partial s}{\partial t}=-2\gamma_{1}[T_{c}-(T-T_{c})\varTheta(T_{c}-T)+2b(\frac{T_{c}}{\alpha})^{2}s].
  \end{equation}
  
It presents a considerable methodological interest to present the same theory simultaneously in terms of internal energy 
and configurational entropy. In accordance with (\ref{b5}) and (\ref{b44}) the expressions for the internal energy and 
its derivatives in this case will look like this
\begin{eqnarray}\label{b46,b47,b48}
u(s)=f^{*}_{0}-(T-T_{c})\varTheta(T-T_{c})s+b(\frac{T_{c}}{\alpha})^{2}s^{2}, \\
T_{rv}=u\prime(s)=f_{0}-(T-T_{c})\varTheta(T-T_{c})+2b(\frac{T_{c}}{\alpha})^{2}s, \\
u\prime\prime(s)=2b(\frac{T_{c}}{\alpha})^{2},
\end{eqnarray}
where
  \begin{equation}\label{b49}
f^{*}_{0}=f_{0}+\frac{\alpha^{2}}{4b}+\frac{\alpha^{2}}{2bT_{c}}(T-T_{c})\varTheta(T_{c}-T),
  \end{equation}
where the dash of variable $s$ is dropped. Here $T_{rv}$ is a current value of the temperature, which in the 
equilibrium state coincides with $T$, that, with the temperature of external thermostat.  
  
\begin{figure}
\includegraphics [width=3.3 in]{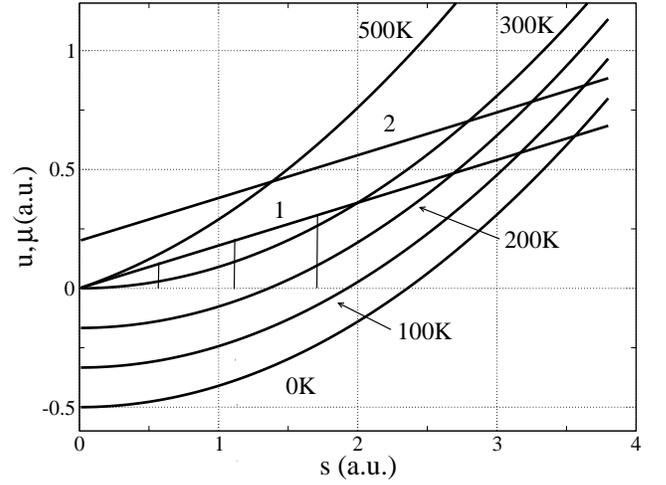}% Here is how to import EPS art
\caption{\label{f9} Dependence of the inernal energy and its derivatives on the entropy}
\end{figure}

The curves of the internal energy and current temperature are resulted in fig. \ref{f9}. 
With the increase of temperature up to critical a minimum of the internal energy grows all of time, remaining at a 
zero value of the configurational entropy. Here the curves of current temperature coincide between itself for all 
temperatures (line 1). Their equilibrium values at different temperatures however differ between itself and equal 
to the temperature of external thermostat. Compare vertical lines in fig. \ref{f9}, drawn between the abscissas axis 
and the straight line of current temperature in the equilibrium states for temperatures $100$, $200$ and $300 K$. 
At a zero temperature of thermostat the equilibrium value of current temperature is equal to the zero.

\begin{figure}
\includegraphics [width=3.3 in]{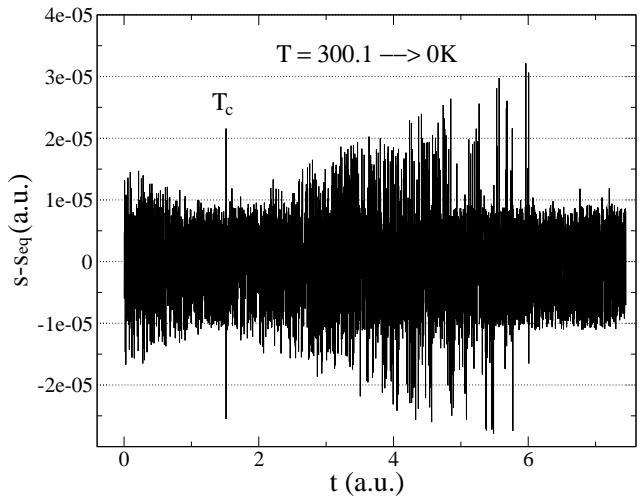}% Here is how to import EPS art
\caption{\label{f10} Dependence of the inernal energy and its derivatives on the entropy}
\end{figure}

At temperatures higher critical ones the minimum of the internal energy moves in the negative unphysical area of 
configurational entropy values. The curve of current temperature with growth of temperature of thermostat begins 
to be evenly moved upwards along the abscise axis. The equilibrium value of current temperature accordingly grows also.

Let us evident that the equilibrium value of current temperature coincides with the temperature of thermostat. 
We define the equilibrium value of the configurational entropy from an obvious condition
  \begin{equation}\label{b50}
\frac{\partial f}{\partial s}=(T-T_{c})\varTheta(T_{c}-T)-T_{c}+2b(\frac{T_{c}}{\alpha})^{2}s=0,
  \end{equation}
from where follows
  \begin{equation}\label{b51}
s_{eq}=\frac{1}{2b}(\frac{\alpha}{T_{c}})^{2}[(T-T_{c})\varTheta(T_{c}-T)+T_{c}].
  \end{equation}
Substituting this value in (47) we get
\begin{eqnarray}\label{b52}
\nonumber
T_{rv}=(T-T_{c})\varTheta(T-T_{c})+\\
+(T-T_{c})\varTheta(T_{c}-T)+T_{c}\equiv T. 
\end{eqnarray}
  
The analogue of the evolution equation (\ref{b20}) in terms of entropy then has more natural form  
  \begin{equation}\label{b53}
\frac{\partial s}{\partial t}=\gamma_{s}(\frac{\partial u}{\partial s}-T).
  \end{equation}  
that, the system is evoluated until a current temperature is accepted the temperature of thermostat.

For a heterogeneous problem equation (\ref{b31}) in terms of configurational entropy it is possible to write down 
in a form
  \begin{equation}\label{b54}
\frac{\partial s}{\partial t}=\gamma_{s}(c\bigtriangleup s+2b(\frac{T_{c}}{\alpha})^{2}s)+f(s).
  \end{equation}

We put this equation in basis for calculation of fluctuations, arising up at passing of critical temperature of 
the system (fig. \ref{f10}). The primary temperature of thermostat got out equal a bit higher critical $T=300.1 K$, 
and goes down slowly. 
  
In the vicinity of critical temperature, predictably, there are intensive long-living fluctuations. It is interest 
that in the strictly critical point long-living fluctuations dies out, and general level of fluctuations go down 
to the thermal background. It is related to that distinction between two types of steady-states in a critical point 
disappears, and they can not serve more by the «traps» of thermal fluctuations for each other. Therefore excrescence 
of fluctuations takes a place not strictly in a critical point, but in some vicinity of it.  
  
Thus, in this paper a theory of the second order phase transitions is considered from four different positions –- 
in terms of free and internal energy in language of order parameter, and also in language of configurational entropy. 
The indicated variants can not simply be taken to each other, and formulation in language of configurational entropy 
seems more preferable. Evolution equation in terms of the internal energy in language of configurational entropy has 
clear physical sense, meaning tendency of current temperature of the system to the temperature of external thermostat. 
The analysis of long-living fluctuations, arising up in the vicinity of critical temperature due to transitions between 
two types of the states, shows that in the strictly critical point the level of fluctuations goes down to the average 
thermal background. 

In conclusion we mark that theory of PT-2, developed here in terms of entropy, can enough correctly describe 
order –- disorder phase transition at transition of Curie point, for example, in a magnetic. Thus heterogeneous terms 
in the free and internal energy describe generation and disappearing of accidental structural fluctuations. At the same 
time, within the framework of this theory it is while problematic to describe structural heterogeneity of the second type, 
namely origin of regular antiphase boundaries \cite{kmy12}. For solution of this problem a complication of the model is required 
by introduction of additional degree of disorder.

%\newpage
%\bibliography{Metlov_MNT}% Produces the bibliography via BibTeX.

\end{document}